\documentclass[aps,prb,reprint,floatfix]{revtex4-1}

\usepackage{graphicx}
\usepackage{amsmath}
\usepackage{amsfonts}
\usepackage{amssymb}

\begin{document}

\title{Optical absorption spectra of finite systems from a conserving Bethe-Salpeter equation approach}

\author{George Pal}
\affiliation{Physikalisch-Technische Bundesanstalt (PTB), Bundesallee 100, 
38116 Braunschweig, Germany}
\author{Yaroslav Pavlyukh}
\affiliation{Institut f\"{u}r Physik, Martin-Luther-University, 
Halle-Wittenberg, Heinrich-Damerow-Strasse 4, 06120 Halle, Germany}
\author{Wolfgang H\"{u}bner} 
\affiliation{Physics Department and Research Center OPTIMAS, 
University of Kaiserslautern, P.O.Box 3049, 67653 Kaiserslautern, Germany}
\author{Hans Christian Schneider}
\email{hcsch@physik.uni-kl.de}
\affiliation{Physics Department and Research Center OPTIMAS, 
University of Kaiserslautern, P.O.Box 3049, 67653 Kaiserslautern, Germany}

\begin{abstract}
We present a method for computing optical absorption spectra
  by means of a Bethe-Salpeter equation approach, which is based on a
  conserving linear response calculation for electron-hole coherences
  in the presence of an external electromagnetic field. This procedure
  allows, in principle, for the determination of the electron-hole
  correlation function self-consistently with the corresponding
  single-particle Green function. We analyze the general approach
  for a ``one-shot'' calculation of the photoabsorption cross section
  of finite systems, and discuss the importance of scattering and
  dephasing contributions in this approach. We apply the
  method to the closed-shell clusters Na$_{4}$, Na$^{+}_{9}$ and
  Na$^{+}_{21}$, treating one active electron per Na atom.
\end{abstract}

\maketitle

\section{Introduction}\label{Sec_Intro}

Optical absorption spectra are a widely used and versatile tool for
the characterization of the electronic properties
of systems of interacting electrons, including their elementary
excitations in condensed-matter systems as well as in molecules and
clusters. The information that can be obtained from absorption and
transmission spectra complements the information accessible from
photoelectron spectroscopy techniques, but the interpretation is more
difficult because transitions in the interacting system are
probed, and the correlated dynamics of electrons and holes in the
many-electron system has to be taken into account in order to compare
with experiments and to identify spectral characteristics.

In a quasiparticle picture, the photoabsorption is determined by the
probability of an electronic transition from an occupied to an empty
(or virtual) quasiparticle state, and can be directly determined from
the electron-hole correlation function. The determination of the
ele-tron-hole correlation function is numerically demanding and
usually effected by time-dependent density functional theory (TD-DFT)
or by solving the Bethe-Salpeter (BS) integral equation derived in the
framework of many-body perturbation theory~\cite{OnidaRevModPhys}.

In the latter approach, the BS equation is often used at the level of a
random-phase approximation (RPA), and a popular choice for the
calculation of the the excited states, which are needed as input for
the BS equation, is the GW approximation for the self-energy, i.e.,
the effective interaction seen by one particle in the presence of the
other carriers. (Here, $G$ is the single-particle Green function and
$W$ the screened interaction.)

In the case of state-of-the-art BS equation
approaches \cite{OnidaPRL,ShirleyPRL,RohlfingLouie1}, which may be
applied to extended and finite systems, the DFT single-particle states
are used as input to GW calculations to obtain the quasiparticle
correction to the Kohn-Sham eigenvalues. Then the equation for the
electron-hole correlation is transformed into an eigenvalue problem,
from which the real parts of the electron-hole transition energies are
obtained.  The numerical calculations needed for these Bethe-Salpeter
approaches are simplified if one assumes an instantaneous, i.e.,
statically screened, electron-hole interaction for the direct part of
the integral kernel of the BS equation. This approach works well with
quite a few extended systemes, such as semiconductors and
insulators. The straightforward inclusion of the full dynamic
screening in the BS equation complicates its numerical solution
tremendously, but it is possible to perform an expansion of the
dynamically screened potential in plasmonic
modes~\cite{RohlfingPRB}. Dynamic screening is also responsible for
excitonic effects even in metals, such as Cu and
Ag~\cite{MariniDelSolePRL}. General attacks on the problem of dynamic
screening in the BS equation go back to the Shindo
approximation~\cite{Shindo}. More recently this problem has been
reexamined in the framework of nonequilibrium Green function theory,
which allows one to derive a Dyson equation for the two-particle
propagator~\cite{BornathKrempSchlanges}.


In addition to standard BS equations there also exist explicitly
time-dependent (nonequilibrium) Green function based methods, which
reduce to the BS equation at the GW-RPA level as a special case, but have
been solved for simpler systems using more complicated two-particle
correlations than those contained in the
GW-RPA~\cite{theDutchGuy,AlmbladhPRL}. It has also been demonstrated
that Green function approaches going beyond the GW-RPA accurately
determine the electronic energy renormalizations and broadening
for collective excitations in extended
systems~\cite{Kwong-Bonitz,KochRMP}.

In finite systems, electronic states and electron-hole transitions
contributing to the absorption spectra are discrete, and are
explicitly treated as such in most Green function based theories as
well as configuration-interaction (CI) and TD-DFT
calculations. However, the experimental signatures due to the discrete
states and transition energies invariably appear with a finite
linewidth in photoemission and absorption measurements, respectively,
due to the finite detector energy-resolution. In particular, the
bunching of levels and transitions in certain energy regions, which
occurs with increasing system size, cannot be resolved experimentally
and may lead to structures that are much broader than the experimental
resolution, so that the underlying contributions from individual
states or transitions cannot be distinguished. This is often accounted
for by adding a phenomenological broadening, which is usually taken to
be the same for all transitions.

The approach of the present paper to the \emph{ab-initio} calculation
of absorption spectra in finite systems uses from the outset a finite
broadening for single-particle states, i.e., finite lifetimes, as it
is rigorously only the case in extended systems. In addition to the
background broadening, however, there result dynamical
interaction-induced features in the absorption spectra, namely
line\-widths and energy shifts, which are closely
connected~\cite{Binder-Koch}. We aim here at the consistent inclusion
of both transition-energy renormalizations and finite linewidths due
to ele-tron-hole correlations for finite systems. We achieve this by
deriving an integral equation from a quantum kinetic description of
the dynamic electron-hole coherences driven by an external, coherent
optical field, including Hartree-Fock (HF) and scattering
contributions. Our approach is equivalent to an equation of the BS
type for the \emph{complex} electron-hole correlation function using
the HF single-par\-ticle energies together with a
\emph{frequency-dependent} integral kernel, which contains correlation
effects beyond the static HF contributions~\cite{Kwong-Bonitz}. The
finite linewidths in our absorption spectra exceed the ``background''
single-particle broadening and thus describe interaction induced
effects including the formation of new peaks. The broadening in our
approach should be distinguished from the artificial broadening in
\emph{nonequilibrium} dynamics~\cite{AlmbladhPRL}.  Since we base our
derivation on the dynamics of electron-hole
coherences~\cite{Kwong-Bonitz} in the interacting system, one can
include self-energy effects and thus make the calculation
self-consistent. 
An important property of the present approach is that it is
straightforward to incorporate the requirement that the (two-particle)
electron-hole correlation function be consistent with the electronic
single-particle properties, so that a conserving approximation (in the
sense of Baym and Kadanoff) results. This approach fulfils important
sum rules by construction~\cite{BaymKadanoff}. For instance, the f-sum
rule for the density-density correlation, which can be read as the
Thomas-Reiche-Kuhn sum rule for the absorption cross section in finite
systems, is directly checked in our numerical calculations. Last, but
not least, we point out that including correlation effects in a
consistent way by examining the equations for electron-hole coherences
makes the interdependence between one- and two-particle properties
clear. This approach therefore avoids possible inconsistencies of a
two-step procedure of calculating first the single-particle
properties, and then solving a BS equation with an effective
two-particle interaction, as it is done in state-of-the-art BS
equation approaches.

Our approach to the \emph{ab-initio} calculation of absorption spectra
in finite systems is based on a recent development of a ``consistent''
GW theory for finite systems~\cite{pal-epjb}, and it has been applied
to extract structural information from cluster absorption
spectra~\cite{pal-jcp}. In the present paper, we are mainly concerned
with a detailed discussion of the general method for computing
absorption spectra. We restrict ourselves to the derivation of a
``one-shot'' calculational method using Hartree-Fock energies and a
constant ``background'' broadening. This is appealing for its
simplicity, and also because it preserves the distinction between HF
and correlation contributions throughout the calculations.  The
present paper expands on the short discussion given in
Ref.~\cite{pal-jcp}, and a few illustrative applications to small
metal clusters, such as Na$_{4}$, Na$^{+}_{9}$ and Na$^{+}_{21}$,
which exhibit rather complicated correlation mechanisms between the
interacting electrons and which develop true collective excitations
with increasing clusters size.


\section{Theory}\label{section:Theory}

In this Section we provide the derivation of the equation for the
electron-hole correlation function, which determines the
photoabsorption cross section. 
We start by summarizing the connection between these quantities. We
then provide the definitions and important properties of the
electron-hole correlation functions in
Sec.~\ref{subsection:QuanumKinetics} using non-equilibrium Green
function theory.  Sec.~\ref{subsec:kinetic-equation} contains the
derivation of the dynamical equation for the electron-hole coherence
in the nonequilibrium formalism, and in Sec.~\ref{subsec:BS} this
result is translated back into a two-particle equation for the
electron-hole correlation function.

The central quantity for the calculation of the optical absorption is
the retarded density-density correlation function
\begin{equation}
\label{chi-rr}
\chi^{\mathrm{r}}(\vec{r},t;\vec{r}',t') =
\langle \rho(\vec{r},t)\rho(\vec{r}',t')\rangle
- \langle \rho(\vec{r},t)\rangle \langle \rho(\vec{r}',t')\rangle
\end{equation}
where $\rho(\vec{r},t)=\psi^{\dag}(\vec{r},t) \psi(\vec{r},t)$ is
the particle density operator, expressed by creation and
annihilation operators $\psi^{\dag}$ and $\psi^{\dag}$,
respectively. 
We use the expansion
\begin{align}
\label{basisChi}
\chi^{\mathrm r}(\vec{r},t; \vec{r'},t') =  \sum_{n_1 \dots
n_4} & \langle
n_{1}n_{3}|\chi^{\mathrm{r}}(t,t^{\prime})|n_{2}n_{4}\rangle\\
\times &
\varphi^*_{n_1} (\vec{r})\varphi_{n_2} (\vec{r}) \varphi^*_{n_3} (\vec{r}')
\varphi_{n_4} (\vec{r}') \nonumber
\end{align}
in a basis of molecular orbitals $\{\varphi_{n}(\vec{r})\}$, where $n$
labels the orbital. The density-density correlation also
describes the linear response of the system to weak perturbation, which is described here
by the coupling of a field, with matrix element in the orbital basis $U_{n_1n_2}(t)$, 
to the system Hamiltonian $H_{\mathrm{sys}}$ via $H(t)=
H_{\mathrm{sys}} + H_{\mathrm{ext}}(t)$ with
\begin{equation}
\label{H-sys}
  H_{\mathrm{ext}}(t) = \sum_{n_1,n_2}\rho_{n_1n_2}(t) U_{n_1n_2}(t).
\end{equation}
The matrix element occurring in
Eq.~\eqref{basisChi} can be expressed as the functional derivative
\begin{equation}
\label{DDcorFun_n}
\langle
n_{2}n_{1}|\chi^{\mathrm{r}}(t,t^{\prime})|n_{3}n_{4}\rangle=\left.
\frac{\delta\langle\rho_{n_{1}n_{2}}(t)\rangle}{\delta U_{n_{3}n_{4}}(t^{\prime}%
  )}\right\vert_{U=0},
\end{equation}
where $\rho_{n_{1}n_{2}}(t)= c_{n_{1}}^{\dag }(t)c_{n_{2}}(t)$ and
$c_{n}$ ($c_{n}^{\dag}$) denotes the annihilation (creation) operator
for an electron in the orbital $\varphi_{n}$. 
Since the external field
couples only electron and hole states, and in linear response the
electronic distribution functions $\rho_{nn}$ are unchanged, the only
nonzero elements in Eq.~\eqref{DDcorFun_n} are those with indices that
pair occupied (electron) and unoccupied (hole) states, i.e.,
Eq.~\eqref{DDcorFun_n} is the electron-hole correlation function.
The photoabsorption cross section thus takes the
form~\cite{GrabowskiGarciaBennemann}
\begin{equation}
\sigma(\omega)=\frac{1}{3} \frac{\omega}{\varepsilon_{0}c}
\sum_{n_{1}\dots
n_{4}}\mathrm{Im}\langle n_{2}n_{1}|\chi^{\mathrm{r}}(\omega)|n_{3}%
n_{4}\rangle ( \vec{d}_{n_{1}n_{2}}\cdot\vec{d}_{n_{3}n_{4}}
), \label{crosssection_n}
\end{equation}
where
\begin{equation}
\vec{d}_{n_{1}n_{2}}=\int\varphi_{n_{1}}^{\ast} (\vec{r})\,
e\vec{r} \, \varphi_{n_{2}}(\vec{r}) d^{3}r
\end{equation}
are the electric dipole
matrix elements in the molecular orbital representation.

\subsection{Density-density correlation function}
\label{subsection:QuanumKinetics}

The problem in the determination of the absorption cross section is
the calculation of $\chi^{\mathrm{r}}(\omega)$, i.e., the Fourier
transformation in $t-t'$ of Eq.~\eqref{DDcorFun_n}.  To obtain
directly an equation for $\chi^{\mathrm{r}}(\omega)$, we take the
detour of deriving dynamical equations for the density response, which
are capable of describing more general non-equilibrium situations, and
adapt these for the response to weak perturbations. This approach is
already implicit in the original treatment of conserving
approximations for the two-particle correlation
function~\cite{BaymKadanoff}, and has recently been exploited
numerically~\cite{Kwong-Bonitz,pal-epjb}. The functional derivative in
the definition of the response function in Eq.~\eqref{DDcorFun_n} can,
in principle, be carried out by solving a \emph{dynamical} equation
for the density~$\langle \rho_{nn'}(t) \rangle$ in the presence of an
external field~$U_{nn'}(t')$. The non-equilibrium dynamics can be
calculated in any many-particle formalism; here we use nonequilibrium
Green functions~\cite{Kwong-Bonitz,BaymKadanoff}, because they offer
the choice to introduce renormalized quasiparticle properties at the
single-particle level. For arbitrary nonequilibrium situations, this
formalism yields \emph{coupled} dynamical equations for kinetic and
spectral Green functions, i.e., $G^<(t_1,t_2)$ and
$G^{\mathrm{r}}(t_1,t_2)$, which depend on two times, even though we
need only the time-diagonal
\begin{equation}
\langle \rho(t) \rangle =-i\hbar G^{<}(t,t)
\end{equation}
for the determination of $\chi^{\mathrm{r}}$.  Calculating the kinetic
and spectral Green functions including their dependence on two times
is numerically extremely demanding, so that it is only feasible for
simple systems~\cite{theDutchGuy,AlmbladhPRL}. Instead of such a
two-time calculation we would like to determine the optical response
of a system at $T =0$\,K or in equilibrium in terms of a quasiparticle
\emph{spectrum}, i.e., we would like introduce an approximation that
replaces the two-time Green functions by
\begin{align}
 G^<(t_1,t_2) \to \mbox{}& G^<(t) \equiv G^{<}(t,t)\\
 G^{\mathrm{r}}(t_1,t_2) \to\mbox{} & G^{\mathrm{r}}(t_1 - t_2)\\
& =
 \frac{1}{2\pi}
\int G^{\mathrm{r}}(\omega) e^{-i\omega(t_1-t_2)}d\omega
\nonumber
\end{align}
In other words, in such an approximation the time dependence $t_1 -
t_2$ essential for the quasiparticle spectrum, is separated from the
time dependence $t = (t_1 + t_2)/2$ essential for the kinetics of the
quasiparticles. This is achieved by the generalized Kadanoff-Baym
ansatz, which reads
\begin{align}\label{GKBA1}
G^{\gtrless}_{n_1n_2}(t_1 ,  t_2)=\mbox{}& i\hbar \,
G_{n_1n_1}^{\mathrm{r}}(t_1-  t_2)\,  G^{\gtrless}_{n_1n_2}( t_2) \\
& \quad - i\hbar\,  G^{\gtrless}_{n_1n_2}(t_1) \,
G_{n_2n_2}^{\mathrm{a}}(t_1- t_2) \ .  \nonumber
\end{align}
We then end up with a kinetic equation for $G^{<}(t)$ that includes
the quasiparticle properties via its dependence on
$G^{\mathrm{r}}(\omega)$, i.e., the Fourier transform of
$G^{\mathrm{r}}$ with respect to $t_1 - t_2$. 

In our recent paper, Ref.~\cite{pal-epjb}, we were interested in the
determination of the electronic quasiparticle properties, i.e.,
renormalized electronic energies and electronic lifetimes, in the GW
approximation, and used different approximations for the
density-density correlation function $\chi$. Here, we summarize the
connection of the quasiparticle properties with $\chi$ and the
screened Coulomb potential. The quasiparticle properties are directly
determined by the interacting Green function
\begin{equation}
\label{G0-G}
 \left[G_0^{-1}(1,\bar{2})-\Sigma(1,\bar{2})\right]G(\bar{2},1') = \delta(1-1'),
\end{equation}
where a space-time integration, or equivalently, a summation over the
single-particle indices and a time integration, for quantities with an
overbar is understood. Eq.~\eqref{G0-G} depends on the
non-interacting Green function $G_0$ (here taken to be the HF Green
function) and the self energy (here taken to be the GW self energy)
\begin{equation}
\label{Sigma}
  \Sigma(1,2) = i\hbar G(1,2) W(2,1).
\end{equation}
(In the orbital representation, Eq.~\eqref{Sigma} is given by
Eq.~\eqref{GW_time} below.)  In terms of the density-density
correlation function, which is equal to the reducible polarization function,
the screened interation~$W$ is an auxiliary quantity given by
\begin{equation}
\label{W-v}
  W(1,1') = v(1,1') 
  + v(1,\bar{2}) \chi(\bar{2},\bar{3}) v(\bar{3},1') .
\end{equation}
Note that the quasiparticle properties depend on the two-particle
correlation function $\chi$ via Eqs.~\eqref{G0-G}--\eqref{W-v}, while
the two-particle correlation function depends on the single-particle
properties via the generalized Kadanoff-Baym an\-satz, so that the
single- and two-particle functions are connected and the full problem
becomes in principle self-consistent.

The following remark on the role of the orbital representation used
above is in order. First of all, since the dipole matrix elements in
the HF representation connect electron states, i.e., unoccupied states
above the LUMO, with hole states, i.e.  occupied states below the
HOMO, the quantities $\langle \rho_{n_{1}n_{2}}(t)\rangle$ in
Eq.~\eqref{DDcorFun_n} describe coherent electron-hole amplitudes in
the presence of an external potential $U$. We will refer to these
expectation values, which are nonzero only in non-equilibrium, as
electron-hole ``coherences.''  The definition of these coherences
depend on the orbital representation. Here, we always use the
representation in HF molecular orbitals.  This will prove particularly
important in the next section, because it allows us to conveniently
separate the HF contribution in the equation for the electron-hole
coherence.

\subsection{Kinetic equation for the density response}
\label{subsec:kinetic-equation}

Our general approach to the calculation of the Green function
$G$ is discussed in some detail in Ref.~\cite{pal-epjb}. Thus we
concentrate here on the derivation of the density-density correlation
function. To calculate the action of the weak external field~$U$, we
start from the equation of motion for $G^<(t_1,t_2)$ evaluated at
equal times $t_1=t_2=t$ which reads
\begin{align}
\label{dtGtt}
 i  \hbar & \frac{\partial}{\partial t}  G^<_{n_1n_2}(t)= 
\\
& \sum_{n_3} \Bigl\{\left[\, T_{n_1n_1}\delta_{n_1n_3}+
U_{n_1n_3}(t)+\Sigma^{\mathrm{HF}}_{n_1n_3}(t) \, \right]
G^<_{n_3n_2}(t,t) \nonumber
\\
&\quad - G^<_{n_1n_3}(t,t) \left[\, T_{n_2n_2}\delta_{n_2n_3}+
U_{n_3n_2}(t)+\Sigma^{\mathrm{HF}}_{n_3n_2}(t) \right]\Bigr\} \nonumber
\\
& \quad + S_{n_1n_2}(t). \nonumber
\end{align}
Here, $T$ is the non-interacting part of the Hamiltonian $H_{\mathrm{sys}} = T+V$,
and includes the core potential in our case.  The scattering
contribution describing correlations beyond the mean-field (HF)
approximation is given by
\begin{align}\label{Scat}
 S_{n_1n_2}(t) = & \sum_{n_3} \int_{-\infty}^{t} d \bar t
\left[ \, \Sigma^>_{n_1n_3}(t, \bar t) \, G^<_{n_3n_2}(\bar t,t)
\right.   \\
& \left.+ G^<_{n_1n_3}(t,\bar t) \, \Sigma^>_{n_3n_2}(\bar
t,t) -(\gtrless \leftrightarrow \lessgtr) \, \right]. \nonumber
\end{align}
Also, the instantaneous HF self-energy
\begin{align}\label{HFselfenergy}
\Sigma^{\mathrm{HF}}_{n_1n_2}(t_1)= &   -i\hbar  \sum_{n_3n_4} \langle
n_1n_3 | v|n_2n_4 \rangle \, G^<_{n_3n_4}(t_1,t_1)  \\
& + i\hbar
\sum_{n_3n_4} \langle n_1n_2 |v|n_3n_4 \rangle \,
G^<_{n_3n_4}(t_1,t_1) \nonumber 
\end{align}
has been separated from the dynamic correlation contribution via the
identity
\begin{align}\label{Sigmaretadv}
\Sigma^{\mathrm r/a}_{n_1n_2} & (t_1,t_2)
 =   \Sigma^{\mathrm{HF}}_{n_1n_2}(t_1) \, \delta(t_1-t_2) \\
& \pm   \Theta  \left(\pm (t_1-t_2)\right) \big[ \Sigma^>_{n_1n_2}(t_1,
t_2)- \Sigma^<_{n_1n_2}(t_1, t_2) \big]. \nonumber
\end{align}
In Eq.~\eqref{HFselfenergy} and below, $v$ is the bare Coulomb matrix element, with the
index structure defined as
\begin{align}
\label{indexCoulomb} \langle n_1n_2 | v |n_3n_4 \rangle  =\int &
d^3r_1 d^3 r_2  \, \varphi^*_{n_1}({\vec r}_1) \varphi^*_{n_2}({\vec
r}_2) \\
& \times v(|\vec{r}_1 -\vec{r}_2|) \, \varphi_{n_3}({\vec r}_1)
\varphi_{n_4}({\vec r}_2) . \nonumber
\end{align}
Eq.~\eqref{Sigmaretadv} is quite general and holds regardless of the
level of approximation for the self-energy, which determines the
functional form of $\Sigma^{\gtrless}$. Due to the separation between
the HF and the dynamic correlation contributions to the self-energy,
and since we are not attempting a self-consistent calculation, it is
natural to use for the basis set functions $\{\varphi_{n}(\vec{r})\}$
the HF molecular orbitals, as opposed to DFT-based approaches that
lump the exchange (Fock) and correlation part together, and separate
out only the direct (Hartree) contribution to the self-energy.

Specifically, we use the GW approximation for the dynamic correlation
contribution
\begin{equation}
\label{GW_time}
\Sigma_{n_1n_2}^{\gtrless}(t, \bar t)=  i\hbar \sum_{n_3n_4}
G^{\gtrless}_{n_3n_4}(t, \bar t) \, \langle n_1n_2 |
W^{\lessgtr}(\bar t, t)| n_3 n_4 \rangle
\end{equation}
where matrix elements of the screened Coulomb interaction $W$ have
been introduced in analogy with Eq.~\eqref{indexCoulomb}. Note that
Eq.~\eqref{GW_time} is just Eq.~\eqref{Sigma} written in the basis of
the HF orbitals.

For the two-time Green functions from the correlation term we employ
the generalized Kadanoff-Baym quasiparticle ansatz~\eqref{GKBA1}.
Note that we have not specified yet which $G^{\mathrm{r}}(t-\bar t)$
we use to describe the spectral properties of the electrons. If one
uses the results of a GW calculation, the calculation becomes self
consistent, because the screened interaction depends on $\chi$. We do
not intend to pursue the self-consistent problem further in the
present paper, so that we neglect the correlation corrections to the
HF single-particle properties in the following. With this
approximation the retarded and advanced Green functions at the HF
level are given by
\begin{equation}
\label{GKBA2}
i\hbar \, G_{n_1n_1}^{\mathrm{r\ (HF)}}(t-\bar t) = \Theta(t-\bar t)
\exp \{-\frac{i}{\hbar}
\tilde \epsilon_{n_1}(t-\bar t)\}.
\end{equation}
For clusters, the corrections to the ground-state energies are
expected to be small so that a Green function constructed from HF
eigenvalues constitutes a reasonable starting point. A non-zero
quasiparticle broadening $\gamma$ ensures the proper behavior of the
HF retarded and advanced Green functions, and we use here and in the
following the notation $\tilde\epsilon_n=
\epsilon^{\mathrm{HF}}_n-i\gamma$. With this approximation,
Eq.~\eqref{dtGtt} is a kinetic equation for the density response,
which has a parametric dependence of the electronic quasiparticle
properties via~Eqs.~\eqref{Scat}--\eqref{GKBA2} and which is closed
in the sense that it does not depend on Green functions with
off-diagonal time arguments.

Before we can actually compute $\chi^{\mathrm{r}}$ as the functional
derivative of Eq.~\eqref{dtGtt} we need a few more steps. The
functional derivative essentially captures the \emph{linear
dependence} of the density on the external potential~$U$. To first
order in $U$, or, equivalently, for a weak external potential, only
density fluctuations, i.e., $\langle \rho_{nn'}(t) \rangle = -i\hbar
G^<_{nn'}(t,t)$ with $n\neq n'$, are influenced by the field,
whereas the distribution functions $f_n(t) = \langle \rho_{nn}(t)
\rangle$ are unchanged and therefore time-independent. Specializing
to the case of such a weak $U$ we can rewrite Eq.~\eqref{dtGtt} as
\begin{align}\label{eqmotGonet}
&\big(i\hbar\frac{\partial}{\partial t}
-\epsilon_{n_{1}}^{\mathrm{HF}}+\epsilon_{n_{2}
}^{\mathrm{HF}}\big)G_{n_{1}n_{2}}^{<}(t)  +(f_{n_{1}}-f_{n_{2}})\Bigl[\frac
{i}{\hbar}U_{n_{1}n_{2}}(t)  \nonumber  \\
&  + \sum_{n_{3}n_{4}}\big(\langle
n_{1}n_{3}|v|n_{2}n_{4}\rangle-\langle
n_{1}n_{2}|v|n_{3}n_{4}\rangle\big)G_{n_{3}n_{4}}^{<}(t)\Bigr] \nonumber \\ 
& \qquad = S_{n_{1}n_{2}}(t) .
\end{align}
Here, we have introduced the HF eigenvalues $\epsilon_{n_{1}}^{\mathrm{HF}}$ and distributions $f_n=
f(\epsilon_n^{\mathrm{HF}})$ where $f(\hbar\omega)$ denotes the
Fermi function, so that $f_n$ is equal to~1 for occupied and~0 for
unoccupied states at $T=0$\,K. In Eq.~\eqref{eqmotGonet}, $S$ stands
for the scattering term  of Eq.~\eqref{Scat}, which includes now the
generalized Kadanoff-Baym ansatz. With the assumption of a weak
external field, we should also consistently linearize the scattering
term with respect to the $G^{\gtrless}_{n\neq n'}$ and replace
factors of $G^<_{n n}(t,t)$ by their equilibrium value, $i\hbar
f(\epsilon_n^{\mathrm{HF}})$.  With these modifications, the
linearized kinetic Eq.~\eqref{eqmotGonet} for the density response
assumes a form, for which we can compute the functional derivative.


\subsection{Bethe-Salpeter equation for $\chi^{\mathrm r}$}
\label{subsec:BS}

The equation for $\chi^{\mathrm{r}}(t,t')$ is obtained by functional
differentiation of Eq.~\eqref{eqmotGonet} with respect to $U(t')$ and
letting $U\to 0$ afterwards. This is done by replacing everywhere
the term $\delta G_{n_1n_2}^<(t)/\delta U_{n_3n_4}(t')$ with
$-i\hbar \langle n_1n_2 |\chi^{\mathrm r}(t-t')| n_3 n_4\rangle$,
cf.~Eq.~\eqref{DDcorFun_n}.  Apart from the $\delta G/\delta U$,
functional differentiation of Eq.~\eqref{eqmotGonet} also yields
contributions of the form $\delta W/\delta U$ but these are
neglected, in agreement with other GW based
approaches~\cite{LouiedWdU}.  The result depends only on $t_1 -
t_2$, and we can Fourier transform in the time difference to obtain
the frequency-dependent $\chi^{\mathrm{r}}(\omega)$ in the form
\begin{align}\label{eqnmotchi_w}
& \big(  \hbar \omega- \epsilon_{n_1}^{\mathrm{HF}}
+\epsilon_{n_2}^{\mathrm{HF}} \big)
\langle n_1 n_2 | \chi^{\mathrm r} (\omega) | n_3 n_4 \rangle   \\
& \qquad + \, (f_{n_1}-f_{n_2}) \Big[ \delta_{n_1n_3} \delta_{n_2n_4} + \nonumber \\
& \sum_{n_5n_6}
\Big( \langle n_1 n_5|v|n_2n_6 \rangle -  \langle n_1 n_2|v|n_5n_6 \rangle
\Big) \langle n_5 n_6 |
\chi^{\mathrm r}(\omega) | n_3 n_4 \rangle \Big] \nonumber \\
&\qquad  +  \sum_{n_5n_6} \langle n_1 n_2 | \Delta (\omega) | n_5 n_6
\rangle \langle n_5 n_6 | \chi^{\mathrm r} (\omega) | n_3 n_4
\rangle = 0 \nonumber
\end{align}
with the correlation kernel
\begin{align}\label{finalD}
\langle  n_1  n_2 & |\Delta(\omega)| n_3n_4\rangle = \nonumber\\
  \sum_{n_5n_6} \Big[ & \frac{ f_{n_1} (1-f_{n_5}) f_{n_6} + (1-
f_{n_1} ) f_{n_5} (1-f_{n_6})}{\hbar \omega -
\epsilon^{\mathrm{HF}}_{n_6} + \epsilon^{\mathrm{HF}}_{n_5} -
\epsilon^{\mathrm{HF}}_{n_1}+ \epsilon^{\mathrm{HF}}_{n_4}
+i(\gamma_{n_1}+\gamma_{n_4}) } \nonumber \\  & \qquad \times \langle n_1n_5 | v |
n_3n_6 \rangle \langle n_6n_2 | v | n_5n_4 \rangle  \nonumber
\\
+   & \frac{  f_{n_2} (1-f_{n_5}) f_{n_6} + (1-
f_{n_2} ) f_{n_5} (1-f_{n_6}) }{ \hbar \omega +
\epsilon^{\mathrm{HF}}_{n_6} - \epsilon^{\mathrm{HF}}_{n_5} -
\epsilon^{\mathrm{HF}}_{n_3}+ \epsilon^{\mathrm{HF}}_{n_2}
+i(\gamma_{n_3}+\gamma_{n_2}) } \nonumber \\
 & \qquad  \times \langle n_1 n_5 | v | n_3 n_6 \rangle \langle n_6
n_2 | v | n_5 n_4 \rangle\nonumber
 \\
- \delta_{n_1n_3}  & \sum_{n_7}  \frac{  f_{n_7}
(1-f_{n_6}) f_{n_5} + (1- f_{n_7} ) f_{n_6} (1-f_{n_5})}{
\hbar
\omega - \epsilon^{\mathrm{HF}}_{n_6} + \epsilon^{\mathrm{HF}}_{n_5}
- \epsilon^{\mathrm{HF}}_{n_3}+ \epsilon^{\mathrm{HF}}_{n_7}
+i(\gamma_{n_3}+\gamma_{n_7}) } \nonumber \\
 & \qquad  \times  \langle n_4n_5 | v | n_7n_6 \rangle \langle n_6n_2 | v | n_5n_7 \rangle\nonumber
 \\
-\delta_{n_2n_4}  & \sum_{ n_7}  \frac{ f_{n_7}
(1-f_{n_6}) f_{n_5} + (1- f_{n_7} ) f_{n_6} (1-f_{n_5})}{
\hbar
\omega + \epsilon^{\mathrm{HF}}_{n_6} - \epsilon^{\mathrm{HF}}_{n_5}
- \epsilon^{\mathrm{HF}}_{n_7}+ \epsilon^{\mathrm{HF}}_{n_4}
+i(\gamma_{n_7}+\gamma_{n_4}) } \nonumber \\
 & \qquad  \times  \langle n_1n_5 | v | n_7n_6 \rangle
\langle n_6n_3 | v | n_5n_7 \rangle   \Big].
\end{align}

The functional form of Eq.~\eqref{eqnmotchi_w} for $\chi^{\mathrm{r}}$
allows one to draw conclusions how many-particle effects will show up
in the calculated absorption spectra. Note first that the direct and
exchange Coulomb contributions renormalize the transition energies
together with the real part of the correlation (or scattering)
contribution~$\Delta$. The imaginary part of~$\Delta$ describes an
interaction-induced dephasing of electron-hole coherences driven by
$U$. Thus the imaginary part of~$\Delta$ determines the line
broadening of the absorption resonances. Note that the complex
quantity~$\Delta$ contains contributions to resonance shifts as well
as the broadening; both effects are therefore related. The
renormalization/broadening contributions~$\Delta$ are determined by a
coupling of each electron-hole transition to all other transitions due
to the Coulomb interaction. Although the single-particle broadening
$\gamma$ (inverse lifetime) of each electron or hole state, or more
generally, the single-particle spectral functions, influence the
broadening of the transitions, Eq.~\eqref{finalD} shows that this
influence is not described by a simple convolution of the electron and
hole lifetimes. Rather, the line broadening for all absorption
resonances arises due to an interaction-induced coupling of individual
electron-hole transitions.  The ``collective'' origin of the damping
should be distinguished from the question whether a given resonance is
due to collective electron-hole excitations, such as excitons or
plasmons, or whether it mainly stems from a single electron-hole
transition.


\section{Results}

In this Section, we apply Eqs.~\eqref{crosssection_n}
and~\eqref{eqnmotchi_w} together with Eq.~\eqref{finalD} to compute the
photoabsorption cross sections for the clusters Na$_{4}$, Na$_{9}^{+}$
and Na$_{21}^{+}$ and we compare our calculations with experimental
results. The photoabsorption cross section of Na$_{4}$, Na$_{9}^{+}$ and
Na$_{21}^{+}$ has been calculated by different approaches, including:
the jellium model
approach~\cite{Ekardt2,Ekardt3,KummelBrackReinhardPRB,BrackRMP,RomanPRB,PachecoPRL},
highly correlated quantum chemistry methods such as multireference
(double) CI~\cite{BonacicChemRev}, a combination of GW and BS equation
approaches~\cite{OnidaNa4}, time-dependent density-functional theory
(TD-DFT) in the local density approximation (LDA)~\cite{Ogut} and
beyond LDA with sophisticated exchange-correlation
functionals~\cite{MarquesCastroRubio}.  Another approach, which has
been applied to calculate the photoabsorption cross section of small
sodium cations (including Na$_9^+$), is the TD-LDA (including
generalized gradient corrections) used along phase-space trajectories
obtained within finite-temperature Born-Oppenheimer local spin density
molecular dynamics~\cite{UziPRL}.

We focus on the influence of correlation effects due to the Coulomb
interaction between the electrons from the outermost electronic shell
of the Na ions at $T=$0\,K, and employ the frozen-phonon
approximation, so that the geometrical structure of the clusters is
fixed at the energetically optimized configuration.  Both core-shell
electrons and electrons from the outermost shell influence the
many-particle states of the interacting electron system and thus the
optical spectra.  However, because the core electron states are
energetically well separated from the outermost shell, these states
are only weakly mixed into the electronic states contributing to
transitions in the energy window of interest, which is a few eVs, and
we expect that the core electrons only introduce an additional fine
structure in the optical spectrum in that window. For the clusters
investigated here, we therefore treat one active electon per Na atom
and use effective core potentials to describe the contribution of the
core electrons. If one is interested in accounting for processes
involving electrons from inner shells a two-particle-hole Tamm-Dancoff
approximation \cite{Schirmer2} or an algebraic diagrammatic
construction \cite{Schirmer4} may be used to describe ionization and
double ionization processes in the inner valence region.

We treat here only closed-shell systems although, in principle, the
theory is applicable to any finite system as the index $n$ refers to a
molecular orbital that can be from an open shell system as well.
Since we are interested in the absorption lineshapes and interaction
induced shifts due to \emph{electronic} correlation effects, we view
the structure optimization of the cluster geomtry, and the necessary
quantum-chemistry only as a tool. A relatively simple approach that
fits well with our method is to obtain the HF eigenfunctions and the
optimization of the underlying cluster geometries from a restricted HF
calculation~\cite{YaroslavHubner} performed with the GAUSSIAN 03
quantum chemistry package~\cite{gaussian03}.  In our calculations, the
$3s^1$ valence electron of each Na atom is explicitly treated using
the \textsc{lanl2dz} basis set: a double zeta basis set of
Gaussian-like atomic orbitals in the (3s3p/2s2p) contraction and
relativistic Los Alamos~\cite{LosAlamos} effective core potential
(ECP) for the core electrons. The structure optimization of the
clusters starts from geometry configurations taken from
Refs.~\cite{KummelBrackReinhardPRB,BonacicChemRev,UziPRL,IshiiPRB}.

For the computation of the correlation contribution $\Delta$ from
Eq.~\eqref{finalD} we use $\gamma$=0.25\,eV for the quasiparticle
energies $\tilde \epsilon = \epsilon - i\gamma$. The value for
$\gamma$ used here is slightly modified from  our earlier GW
results, where we obtained for energetically low occupied states
0.085\,eV for Na$_9^{+}$ and 0.29\,eV for Na$_{21}^+$~\cite{pal-epjb}.
However, this choice simplifies the numerical calculation because it
makes it easier to carry out the sums over complicated energy
denominators, and makes $\Delta(\omega)$ a smoother function of
$\omega$. We stress that introducing a fixed \emph{quasiparticle}
broadening for the calculation of $\Delta(\omega)$ does not mean
that we are fixing the broadening of the resonances
in~$\chi^{\mathrm r}$ to this value. Rather, the resonance
broadening, which is responsible for the finite width of the peaks
in the absorption spectrum, is due to the imaginary part and the
complicated index dependence of the full correlation
kernel~$\Delta(\omega)$.

\begin{figure}[tb!]
\includegraphics[scale=.22]{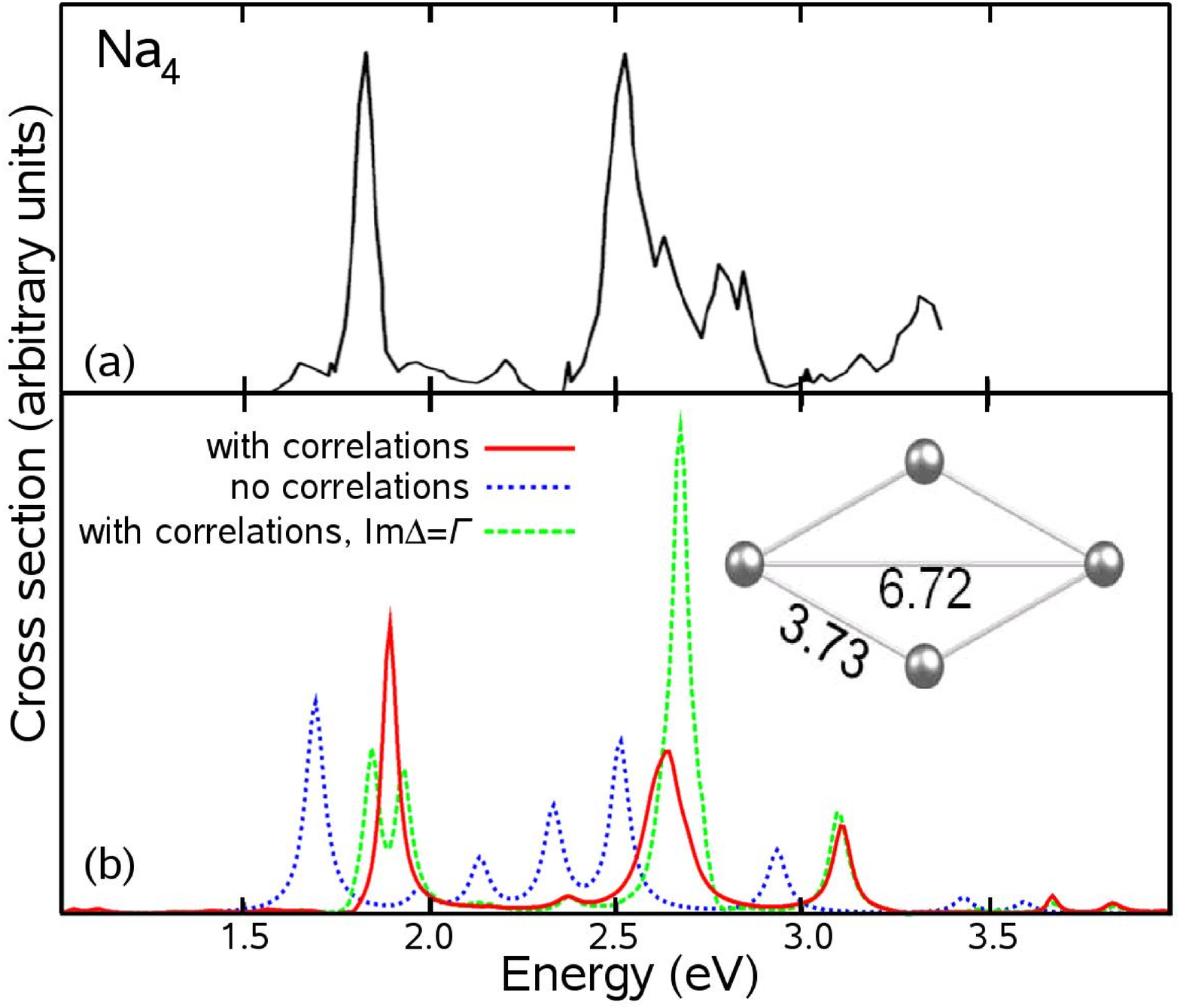}  
\caption{Measured (a) and computed (b) absorption cross section for
  Na$_4$. The experimental result is adapted from
  Ref.~\cite{Na4EXP}. The theoretical results are obtained with (solid
  line) and without (dotted line) correlation contributions in the
  equation for~$\chi^{\rm r}$. The spectrum obtained using only the
  real part (plus a constant imaginary part) of the correlation term
  is also shown (dashed line).  The inset shows the cluster geometry
  with distances in~\AA{}.
\label{fig:Na4}}
\end{figure}

Figure~\ref{fig:Na4} shows the computed absorption cross section for
Na$_4$ together with the experimental result from
Ref.~\cite{Na4EXP}. The main measured peaks which occur around
1.8 and 2.5\,eV are well resolved by the theoretical spectrum when
$\chi^{\rm r}$ is calculated by Eqs.~\eqref{eqnmotchi_w}
and~\eqref{finalD}. To emphasize the importance of the correlation
contribution, we compare this with the cross section calculated
using a constant, phenomenological broadening for each transition.
Technically, this is achieved by using Eq.~\eqref{eqnmotchi_w} with
a constant, diagonal broadening, i.e.,
\begin{equation}
\langle n_1 n_2 | \Delta
(\omega) | n_5 n_6 \rangle=i\delta_{n_1n_5} \delta_{n_2n_6} \Gamma .
\end{equation}
where we have assumed $\Gamma=0.03$\,eV. Using the phenomenological
broadening, the absorption cross section does not compare well with
the experimental data as regards the position and the shape of the
lines. Only the inclusion of the full correlation term yields good
agreement between the calculated and the measured spectra. This result
illustrates that the shift of the absorption resonances due to the
real part of correlation contributions can be on the order of several
tenths of eVs and this shift is needed to obtain agreement with
experiment. In addition to the shift, the imaginary part of $\Delta$,
which is responsible for the broadening, redistributes spectral weight
between the different peaks and leads to an overall very different
shape of the absorption spectrum for the two cases.  To disentangle
the effect of the broadening on the redistribution of the spectral
weight, we have performed a calculation with only the real part of
$\Delta$, plus a diagonal broadening. This yields a splitting of the
peak at 1.8\,eV into two new peaks.  Also, the peak at 2.5\,eV is
blue-shifted and now contains most of the spectral weight.  This is in
agreement with results for extended system, where it is known that the
broadening and level shifts in this type of complex BS equation are
closely connected and cannot be separated by taking into account the
real or imaginary part only.

\begin{figure}[tb!]
\includegraphics[scale=.335]{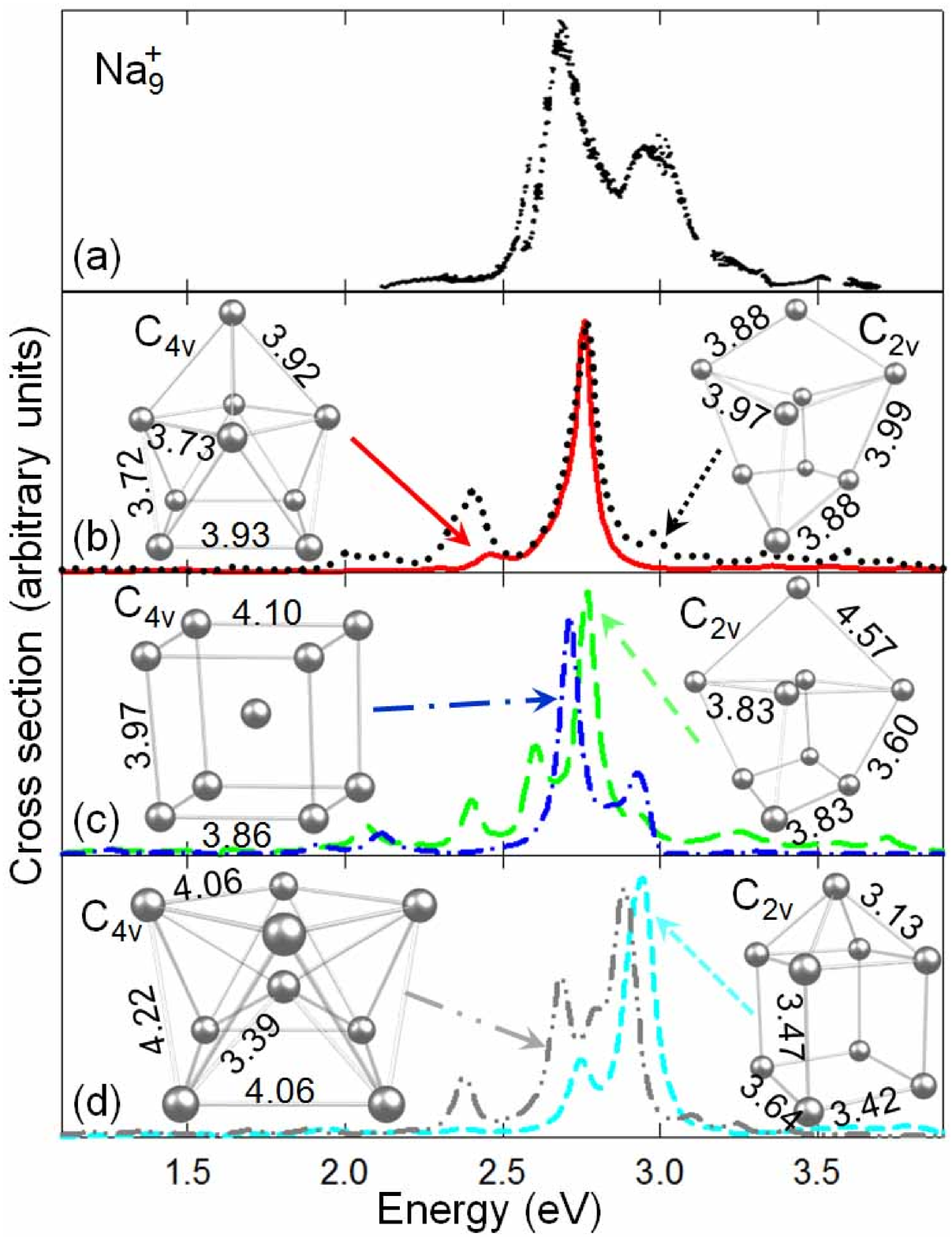} 
\caption{Measured (a) and calculated absorption cross
  sections for Na$_{9}^{+}$. The experimental results are adapted from
  Ref.~\cite{Na921EXP}. The spectra (b)--(d) are calculated for
  six different cluster geometries.  The insets show the cluster
  geometries with distances in~\AA{}. 
\label{fig:Na9}}
\end{figure}

When comparing experimental and theoretical results one should keep in
mind that the geometry configuration, which we obtain by structural
optimization, is important for the calculation, but not well known for
the clusters studied in the experiment. Moreover, finite temperature
and inhomogeneous broadening effects in experiments wash out the
intrinsic spectral features.  To account for the finite temperature
effects and for the unknown cluster geometries which might occur in
experiment, we have considered six different configurations for
Na$_{9}^{+}$ with low ground state energy: three structures with
C$_{\mathrm{4v}}$ symmetry and three with C$_{\mathrm{2v}}$ symmetry,
see Fig.~\ref{fig:Na9}. The total ground state energies of the
clusters differ by less than 0.3\,eV. The comparison between the
measured (from Ref.~\cite{Na921EXP}) and the calculated spectra is
shown in Fig.~\ref{fig:Na9}. The experimental spectra exhibit a
splitting of the main absorption line into a larger peak centered
around 2.68\,eV and a smaller peak around 2.98\,eV. For all the
clusters considered here, the main absorption lines occur in the same
energy interval as the measured lines, but the shape of the peaks does
not compare as well with experiment as in the case of Na$_4$. Both the
C$_{\rm{4v}}$ and the C$_{\rm{2v}}$ clusters from panel (b) of
Fig.~\ref{fig:Na9} yield a single large peak at 2.74~eV. The
C$_{\rm{4v}}$ cluster from panel (c) reproduces best the experiment
with respect to the position and the shape of the absorption
lines. Also, the C$_{\rm{2v}}$ cluster from panel (d) of
Fig.~\ref{fig:Na9} is capable of resolving the smaller experimental
peak at 2.98~eV, but the larger experimental peak at 2.68~eV is not
present in the theoretical spectrum. In an experiment, clusters with
different configurations may contribute to the shape of the absorption
spectrum, but the weight of the contributions due to different
possible configurations cannot be determined from experiment nor from
a $T=0$\,K theory. Given the potential problems for experiment-theory
comparisons, we find a good overall agreement between the measured and
the calculated spectra of Na$_9^+$.

\begin{figure}[t]
\includegraphics[scale=.23]{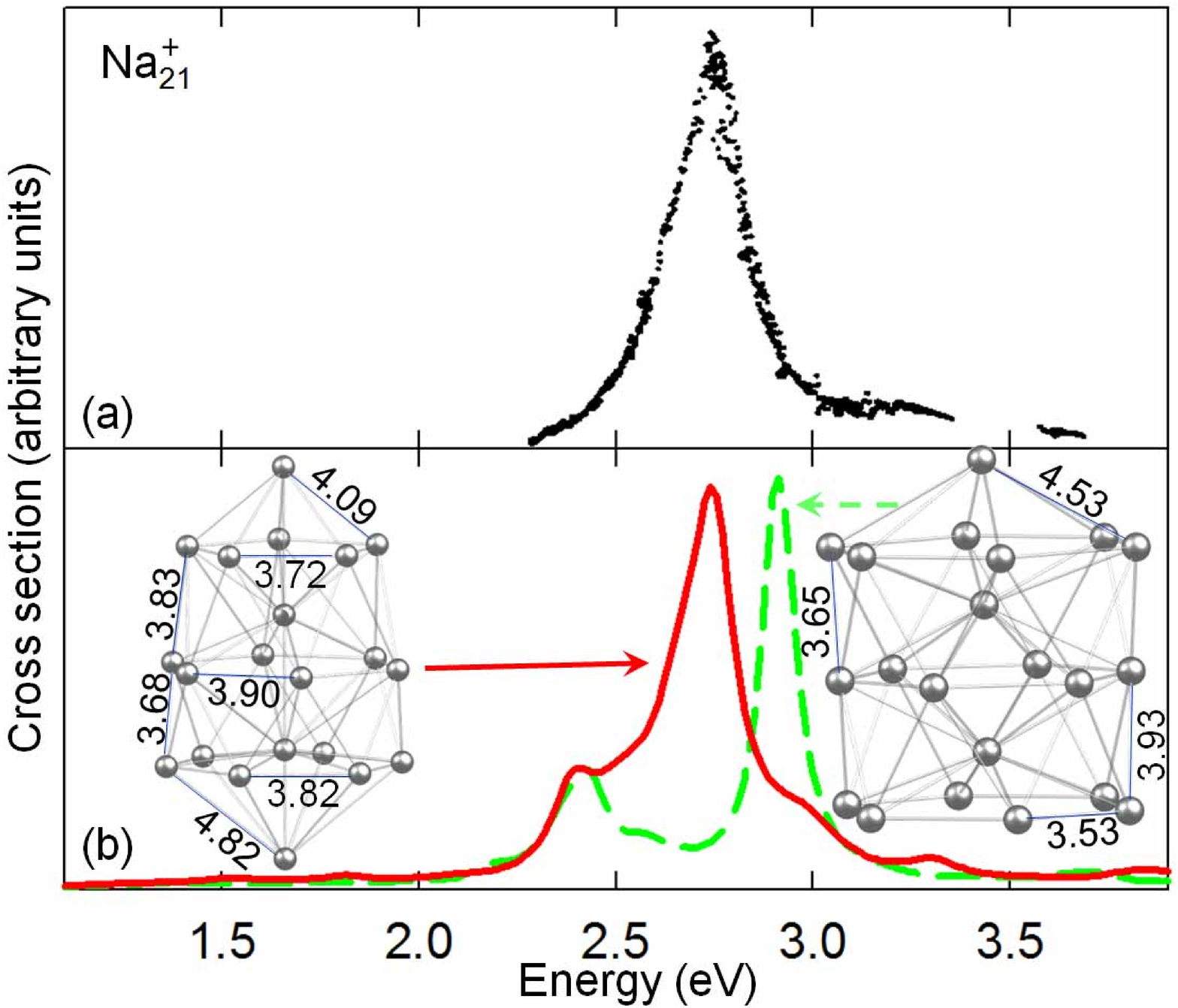} \label{fig:Na21}
\caption{Measured (a) and computed (b) absorption cross section for
  Na$_{21}^{+}$. The experimental result is adapted from
  Ref.~\cite{Na921EXP}. The theoretical spectra are calculated
  for the prolate cluster (solid line), and the structure with
  C$_{\rm{6v}}$ symmetry (dashed line).  The insets show the cluster
  geometries with distances in~\AA{}. }
\end{figure}

Figure~\ref{fig:Na21} shows the comparison between the experimental
and the theoretical cross section of Na$_{21}^{+}$. The measured
spectrum displays a large absorption line centered around
2.74\,eV~\cite{Na921EXP}. Two different cluster geometries were used
in the calculation: one corresponding to a prolate (i.e., elongated)
cluster and one structure with C$_{\rm{6v}}$ symmetry. The
qualitative agreement between theory and experiment is very good
with respect to the position and the shape of the peak for the
prolate Na$_{21}^{+}$. For the cluster with C$_{\rm{6v}}$ symmetry
the agreement is not that good, although the main peaks are situated
in the experimental 2.3--3.2\,eV energy interval.

The different quality of the agreement between theory and experiment
for the Na$_4$ and Na$_{21}^+$ clusters is also an indication that
cluster size plays an important role: For small clusters, the
geometrical configuration is relatively well defined, and the
photoabsorption cross section is characterized by a few discrete
resonances. In the other extreme, when the system contains a large
number of atoms there are many more overlapping individual electron-hole
transitions, so that the resonances are not well
resolved. Moreover, there may be different geometrical
configurations with similar total energies, which may contribute to
the experimental spectrum. For Na$_9^+$, we are neither in the limit
of a small nor a larger cluster so that there is a competition
between the two trends: the spectral density of the single
electron-hole transitions is not yet large enough as to yield
featureless broadened peaks, but the number of the possible
configurations with different electronic properties is large enough
so that so that the cluster geometry present in the experiment is
difficult to predict.

Finally we wish to point out that we have checked numerically that
all the computed spectra shown fulfill the f-sum rule for finite
systems to better than 99\%
\begin{align}\label{fsumrule}
\frac{1}{3}\int  d \omega \, \omega & \sum_{n_1\dots n_4}
( \vec{d}_{n_{1}n_{2}}\cdot\vec{d}_{n_{3}n_{4}} ) \\
& \times {\rm Im} \langle n_2n_1|\chi^{\rm r}(\omega)|n_3n_4
\rangle= -\frac{\hbar^2 \pi}{m}N_e,  \nonumber
\end{align}
where $m$ is the electron mass and $N_e$ is the total number of
electrons in the system. Eq.~\eqref{fsumrule} is the
Thomas-Reiche-Kuhn sum rule for the absorption cross section and is
a measure for the overall quality of the absorption spectrum.

\section{Conclusions}

We have presented an approach to compute the absorption spectra for
finite systems based on a linear response theory for the dynamical
electron-hole coherences in the presence of an external field. Using a
quasiparticle ansatz in the quantum kinetic equation for the
electron-hole coherence, which includes HF and scattering
contributions, allows us to derive a BS equation for the electron-hole
correlation function with a complex, frequency-dependent kernel. The latter
is determined by the scattering (or correlation) contributions to the
electron-hole coherence dynamics. One thus obtains absorption spectra
that include level-shifts of transition energies and a broadening of
the absorption peaks due to electronic interactions. The correlation
induced shifts and the broadening are closely related since they are
the result of the complex, frequency dependent kernel, and the
broadening mimics the bunching of discrete levels in finite
systems. We apply the method to three important Na clusters: Na$_4$ and
the magic number clusters Na$_9^+$ and Na$_{21}^+$, and compare both
the calculated resonance energies and the shape of the absorption
lines with experimental results.

G.P. and W.H. acknowledge support through the
Priority Programme 1153 of the German Research Foundation.

\end{document}